\documentclass{aastex}
\usepackage{graphics}
\usepackage{emulateapj5}
\usepackage{epsf}
\usepackage{color}

\begin{document}

%{\LARGE {VERSION 16 Dec 2015}}

\title{NGC 5195 in M51: Feedback `Burps' after a Massive Meal?}

\author{E. M. Schlegel\altaffilmark{1,4}, C. Jones\altaffilmark{2}, M. Machacek\altaffilmark{2}, L. D. Vega\altaffilmark{3}}

\altaffiltext{1}{Department of Physics and Astronomy, University of
Texas-San Antonio, San Antonio, TX 78249; eric.schlegel@utsa.edu}

\altaffiltext{2}{Harvard-Smithsonian Center for Astrophysics}

\altaffiltext{3}{Fisk University/Vanderbilt University Bridge Program,
Nashville, TN}

\altaffiltext{4}{Vaughan Family Professor}

%\submitted{submitted: 2015 Nov 24}

\begin{abstract}

We describe a double-arc-like X-ray structure lying ${\sim}15-30''$
(${\sim}$0.8-1.7 kpc) south of the NGC 5195 nucleus visible in the
merged exposures of long {\it Chandra} pointings of M51.  The
curvature and orientation of the arcs argues for a nuclear origin.
The arcs are radially separated by ${\sim}15'' ({\sim}1$ kpc), but are
rotated relative to each other by ${\sim}30^{\circ}$.  From an
archival image, we find a slender H${\alpha}$-emitting region just
outside the outer edge of the outer X-ray arc, suggesting that the
X-ray-emitting gas plowed up and displaced the H${\alpha}$-emitting
material from the galaxy core.  Star formation may have commenced in
that arc.  H${\alpha}$ emission is present at the inner arc, but
appears more complex in structure.  In contrast to an explosion
expected to be azimuthally symmetric, the X-ray arcs suggest a focused
outflow.  We interpret the arcs as episodic outbursts from the central
super-massive black hole (SMBH).  We conclude that NGC 5195 represents
the nearest galaxy exhibiting on-going, large-scale outflows of gas,
in particular, two episodes of a focused outburst of the SMBH. 
The arcs represent a clear demonstration of feedback.
\end{abstract}


\section{Introduction}

Through both observations and simulations, significant progress has
been made in understanding the role of supermassive black holes in
galaxy evolution.  It is now widely accepted that at high redshifts,
galaxies undergo an active period of gas cooling, star formation and
rapid black hole growth through radiatively efficient accretion.  This
quasar phase is generally followed by radiatively inefficient
accretion of material onto the SMBH, with the energy from the SMBH
primarily released mechanically through radio-emitting jets that can
inflate bubbles in the hot atmospheres that surround massive
elliptical galaxies (\citealt{Churazov05}; for recent reviews, see
\cite{Fabian2012}, hereafter F12; or \cite{KP2015}, hereafter
KP2015). While many examples of AGN outbursts from SMBHs in massive
elliptical galaxies have been found from Chandra and XMM-Newton
observations, in this paper we present the first evidence for episodic
outbursts from the central black hole in the nearby (8 Mpc),
relatively low-mass early-type galaxy NGC~5195.  NGC~5195 (M51B),
along with the grand-design spiral NGC~5194 (M51A), form the Messier
51 interacting system.

Much of the attention given to M51 focuses on the spiral NGC 5194.
The less-photogenic companion NGC 5195 is variously typed as an SB0-1,
Irregular, or LINER (e.g., NED.ipac.caltech.edu, \citealt{RC3}).  The
nucleus of NGC 5195 is detected in every wavelength from X-rays to
radio.  NGC 5195 tends to receive less attention because of the
significant dust component that is sufficiently dense to virtually
hide the nature of the galaxy \citep{VanDyk1987}.  

Here we describe an {\it X-ray-bright}, double-arc extended structure
lying ${\sim}15-30''$ south of the nucleus of NGC 5195 that is visible
in deep {\it Chandra} observations of M51 (PI K. Kuntz).  We interpret
the double-arc structure to originate from episodic outflows which
have `snow-plowed' H${\alpha}$-emitting material, from the galaxy's
core.  Given the cosmologically recent interaction of NGC 5195 and NGC
5194, estimated to have occurred 50-100 Myr ago \citep{SL2000}, the
structure immediately suggests a reaction to the forced feeding of the
SMBH in NGC 5195.

Estimates of the distance to M51 range from 7.9 Mpc, derived from the
expanding photospheres method for core-collapse supernovae
\citep{Bose2014}, to 8.4 Mpc, based on the planetary nebula luminosity
function \citep{Feldmeier1997}.  We adopt a distance to M51 of 8 Mpc.

\section{Data and Analysis}

\subsection{Data Preparation}\label{DataPrep}

The {\it Chandra} Advanced CCD Imaging Spectrometer (ACIS)
observations of M51 used in this study were obtained from the Chandra
archive and are listed in Table~\ref{m51_obs}.  Seven long pointings
were obtained in Faint mode and total $\sim$760.2 ksec.  Three
observations, totaling 84 ksec, were obtained in Very Faint mode and
are discussed in \S\ref{pics} below.  We do not analyze two
observations: a short 2-ksec observation obtained in Faint mode (ObsID
414) and a ${\sim}$10 ksec observation (ObsID 12562) obtained in Very
Faint mode in which NGC 5195 falls in the gap between the ACIS-I and
ACIS-S CCDs.  These observations are not included in our analysis as
they add little or no signal.

Data preparations were all carried out using the Chandra Interactive
Analysis of Observations (CIAO) software package and followed the
recommended processing.  Each ObsID was examined for times of high
particle background.  Although no such periods were found, short time
intervals were trimmed at the start and end to eliminate possible
residuals; the trimmed times sum to $<$0.1\% of the total exposure.
Gain calibrations were checked and found to match.  The data were
energy-filtered to include only events between 0.4 and 8 keV.  Given
the multiple pointings, the absolute positions of a given ObsID may
differ slightly from the others.  Consequently, we arbitrarily adopted
ObsID 13814 (the longest exposure) as our reference.  The CIAO task
{\tt merge\_obs} was used to place all frames into a common reference,
construct exposure maps for each observation, and merge the exposure
maps and event files to build a flux image.  

Point sources above a flux of ${\approx}5{\times}10^{-16}$ erg
s$^{-1}$ cm$^{-2}$ (corresponding luminosity at the distance of M51 is
${\approx}4{\times}10^{36}$ erg s$^{-1}$) were filtered out from all
subsequent analyses.  {\bf This limit means that the seven bright
sources seen in the field are all filtered from the data.  Five of the
sources are too distant to have any impact on the spectra or profiles
of the arcs.  A 6th source lies too far west of the western edge of
the inner arc to have any impact -- the source is ${\sim}$15'' W of the
edge while the 90\% encircled energy radius at its position is
${\sim}$4''.  The 7th point source sits just S of the nucleus on the
east edge of the inner arc.  Again, the 90\% encircled energy radius
of ${\approx}$4 arcsec, while contaminating some of the arc emission,
remains small relative to the total size of the inner arc.
Consequently, the western half of the arc is uncontaminated by point
source emission.}

\subsection{X-ray Imaging}\label{pics}

\begin{figure} 
\centering 
\caption{(a) {\it Chandra} images of NGC 5195 from the merged long
observations of M51 (PI K. Kuntz).  North is up and east is left for
all frames; the scale bar on the right side of the figure is 1 arcmin
in length.  The nuclei of NGC 5194 (A) and NGC 5195 (B) are labeled, as
is an X-ray-emitting spiral arm lying midway between the nuclei
\citep{Vega2016}.  The image shows events between 0.4 and 8 keV,
binned by a factor of 2, exposure-corrected but without background
subtraction.}
\label{n5195_structure}
\scalebox{0.6}{\includegraphics{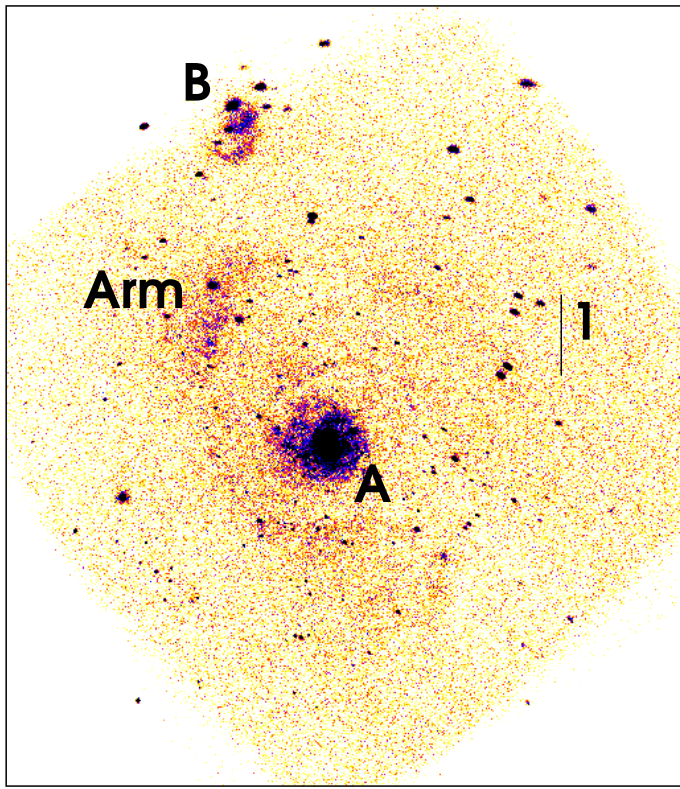}}
\end{figure}

The overall field is shown in Figure~\ref{n5195_structure}(a).  The
nuclei of NGC 5194 and NGC 5195 are labeled `A' and `B,' respectively.
NGC 5194 is nearly centered in the field, so the majority of the galaxy
is covered by the image.  Also visible is the X-ray emission from a
prominent spiral arm lying between the nuclei (discussed in
\cite{Vega2016}).

The nucleus of NGC 5195 lies just inside the ACIS field of the merged
F-mode exposures (Figure~\ref{n5195_structure}).  In all subsequent
discussion, the two X-ray arcs are labeled as `inner,' to describe the
one closest to the nucleus of NGC 5195, and `outer'.  Point sources at
the east end of both the inner and outer arcs are removed from the
analysis of the spectrum and flux of the respective arc as mentioned
previously.

The arcs are separated from the nucleus by ${\sim}$780 (${\sim}15''$)
and 1700 pc (${\sim}30''$), respectively (Table~\ref{arc_pos}).  The
two arcs have broadly comparable curvature: the inner arc spans
$\sim$50-60 degrees (position angles ${\sim}$180-240 degrees measured
Eastward from North), while the outer arc spans ${\sim}$40-50 degrees
(position angles ${\sim}$160-210 degrees).  The curvatures are
consistent with an origin at or near the SMBH.  The outer arc is
approximately centered on a line joining the nuclei of NGC 5195 and
NGC 5194, but the inner arc is centered on a line rotated ${\sim}$30
degrees counter-clockwise from the NGC 5194/5195 line.  Note also that
the inner arc is brighter than the outer arc by about a factor of two
(\S\ref{arcspectra}), consistent with expanding gas.  The curvature,
coupled with the inner {\it vs} outer brightnesses, suggests a nuclear
outburst or similar expulsion of matter with the emission becoming
fainter as the gas expands.  The presence of the two arcs suggests two
episodes of SMBH-driven outflows.  We discuss other possibilities
to explain this structure later in this section.

While we can measure the length and width of each arc, the thickness
$thick$ is unknown.  We adopt a thickness of 100 $thick_{100}$ pc at
the front edge of the outer arc -- a specific thickness at the outer
edge leads to a constant opening angle.  If the arcs were a very broad
angular hemispheric `fan', significant X-ray emission would exist
between the arc and the nucleus, but we do not detect such emission.
Instead, we detect emission that is length- and width-restricted.  The
width restriction implies an episode; the length restriction implies
an angular constriction in a flow with unknown depth or thickness.  An
explosion is expected to be azimuthally symmetric, but the arcs to the
south are not, appearing instead to be two episodes of a `focused'
outflow.

The 85-ksec ACIS VF mode data are shown in
Figure~\ref{n5195_structure}(c).  The data provide a glimpse north of
the NGC 5195 nucleus; a `glimpse' because the signal-to-noise in the
ACIS VF observation is low.  X-ray structure exists to the North,
however, it does not appear to be arc-like, but instead V- or
fan-shaped with the point of the V toward the SMBH.  {\it If} the
density to the north decreases sufficiently quickly, a bubble expanding
northward could open, possibly leading to a V- or fan-shaped
structure.

Are there alternative explanations at this point?  One could interpret
the structure as weak point source(s) plus diffuse emission.  However,
this structure is not consistent with blended point sources, even for
the {\it Chandra} point spread function ${\sim}4'$ off-axis.  We note,
however, that our PSF analysis applies only to sources above
${\approx}100$ counts as fainter ones are inseparable from the diffuse
emission at this off-axis angle.  A definitive analysis requires an
on-axis observation.  {\bf Another possibility is induced star
formation within the arcs; we discuss this interpretation in \S2.5
after presenting observations in other wavelengths.}

A similarly-shaped, fan-like structure also appears to the north in a
continuum-subtracted H${\alpha}$ image of NGC 5195 presented in
\cite{Kaisin2008} and the H${\alpha}$ image from \cite{Hoopes2001}.
We discuss the H${\alpha}$ image in \S\ref{OtherBands}.

\subsection{Surface Brightness Profiles of the X-ray Arcs}

Figure~\ref{RadProf} displays the surface brightness profiles across
both X-ray arcs.  Each profile was built by extracting counts using
an azimuthally-constrained annular region similar to that shown in
Figure~\ref{n5195_structure}(b), but with a larger number of annuli.

The profiles were extracted by matching angular annuli to the shape of
the outer edge, then dividing the region interior to that edge into 20
segments identically shaped.  Leaving aside the detector background,
there are two source backgrounds to consider: the real cosmic
background well-away from M51, and the diffuse background of
NGC 5195.  The counts in radial bins lying outside the outer
edges of the arcs represent the galaxy background.  The inner arc
borders closely on the outer arc.  Consequently the background radial
bin of the inner arc is shorter to avoid overlap with the outer arc.
Finally, the cosmic background was adopted as a large region WSW of
NGC 5195 and sufficiently separated (${\sim}$2.5 arcmin) to avoid any
contamination from point sources or diffuse emission from the galaxy.

The surface brightness profiles are similar in shape: relatively flat
as a function of radius out to the outside edge, then dropping quickly
to the level of the ISM.  That behavior describes a shock.  The large
off-axis angle means the point spread function is broad at this
location (50\% encircled energy width of the PSF ${\sim}$2-2.5 arcsec;
{\it Chandra} Proposers Guide).  Consequently, the arcs are blurred;
to ascertain their shape more accurately will require an on-axis
observation.

\subsection{Spectral Fitting}

{\bf We extract spectra from the two arcs as well as the nucleus for
the purpose of comparing their temperatures and absorption
properties.}

\subsubsection{Overall Spectral Fitting}\label{overall}

We adopted the CCD S7 blank-sky background reprojected onto the
reference frame of NGC 5195.  We used the blank background because (i)
M51 essentially covers the back-illuminated CCD S7, reducing the area
available for an on-chip background; and (ii) possible Si fluorescence
within the instrument must be assessed to determine whether Si
emission is present in the source spectra.  We followed the
CXC-defined analysis thread to reproject that background into the
reference frame of the merged observation, a procedure based on
\cite{Hickox2006}.  We also matched the particle backgrounds, using
events in the 10-12 keV range, of the M51 and blank-sky observations
as described in the Hickox \& Markevitch analysis.  Finally, we
verified the normalization by extracting a background spectrum from
the corners of CCD S7.  The corner spectrum has lower signal-to-noise,
but does verify that the reprojected data match the observed
background.

The spectra for the overall diffuse emission and the backgrounds were
extracted using the regions shown in Figure~\ref{n5195_structure}(b).
The extraction region for the nucleus was a circle centered on the
nucleus of radius 9 arc secs.  Detected point sources were
removed from the source regions using excluded regions corresponding
to $>$97\% of the on-axis point spread function ({\it Chandra}
Proposer's Guide).  Response matrices and effective area files were
built for each source and background region.  The data were binned to
15 counts per bin leading to ${\approx}$225 channels in the diffuse
spectra, of which ${\approx}$80 cover the 0.4-2 keV band, and
${\approx}$480 channels in the nuclear spectrum of which
${\approx}$180 cover the 0.4-2 keV band.

The source + background and background spectra were fit
simultaneously, ensuring the best measure of the flux for the source.
We used XSPEC package v12.8 \citep{Arnaud1996} and the C statistic
\citep{Cash79}.  We adopted a power law for the background model with
several gaussians to mimic small fluorescence features in the
spectrum.  Once the model parameters were fit, we fixed all except the
normalizations for a simultaneous fit of the source + background and
background spectra.  We describe the nuclear spectrum first because we
must refer to it when discussing the spectra of the arcs.

\subsubsection{Nucleus Spectrum and Luminosity}\label{NucSpec}

For the nucleus, we adopted an absorbed power law typical of AGN and
included an optically-thin gas model (described in \S\ref{arcspectra}
below) because our extraction region includes some surrounding gas.
We tested the significance of the gas component: a pure power law model
significantly fails to fit the data, while the combined components do.

The column density was a free parameter with a fitted value of
${\sim}1.5{\times}10^{21}$ cm$^{-2}$.  This value is in close
agreement with the value implied by the column density of K~I measured
by \cite{Ritchey2015}.  The observed luminosity of the nucleus in the
0.5-2 keV band is ${\sim}3.3{\times}10^{38}$ ergs s$^{-1}$ (unabsorbed
L$_X$).  This value indicates that the NGC 5195 SMBH must be in a
radiatively inefficient state (see \S\ref{disc}).

\subsubsection{X-ray Arc Spectra and Luminosities}\label{arcspectra}

For the X-ray arc spectra, we adopted an absorbed, optically-thin
thermal gas model with variable abundances ({\tt vapec} in XSPEC).
{\bf We adopted a power-law component to fit any residual nuclear or
un-removed X-ray binary emission in the arc spectra.  We also adopted
a fixed value for the index (1.6; \citealt{Gilfanov2004}).  The resulting
fit yields an upper limit on the presence of a power law component in
both arcs.}  All of the fitted model parameters are listed in
Table~\ref{fit_table1}.  Initial, fully-unconstrained fits to the
spectra showed repeated consistency for the arcs' temperatures,
normalizations, and abundance values, but not for the column
densities.

We initially fixed the column density at a value of
2.1${\times}10^{20}$ cm$^{-2}$, the Galactic column in the direction
of NGC 5195 from \cite{Schlafly2011}.  {\bf We did so because an
unconstrained fit did not converge on a fixed value of the column.
Once we reached a 'best-fit' state, we also explored fits with the
column densities of the arcs fixed between the Galactic column and the
nuclear value.  The temperatures, abundances, and model normalizations
are consistent with {\it any} value between the known Galactic column
and the fitted nuclear column (0.2 to 1.6${\times}10^{21}$ cm$^{-2}$).
To undertake a proper fit for the column density will require either
(i) additional X-ray data; (ii) an on-axis observation; or (iii)
additional constraints (such as a measured value for the reddening,
E$_{\it B - V}$).}

The initial fits were also done for temperature and normalization with
abundances fixed at solar (= 1.0).  Then each abundance in turn was
fit.  Abundance values that were consistent with 1.0 were fixed at
1.0.  Once the abundances were determined, a final fit was done with
temperature, normalization, and any non-solar abundances as free
parameters.  Errors at the 90\% level on each parameter were then
determined.   

We found the gas temperatures for the two arcs differ, with the outer
arc cooler ($0.41^{+0.06}_{-0.03}$ keV) than the inner arc
($0.65^{+0.04}_{-0.03}$ keV) (Figure~\ref{n5195_spectra}) -- the outer
arc has had a longer time to expand and to cool, as well as plowing
into more material that enhances the cooling.  The fitted model
normalizations also support one's visual impression that the outer arc
is fainter -- they differ by a factor of ${\approx}$1.5 with the outer
arc having a smaller normalization.  The 0.5-2 keV luminosities
calculated from the fitted models are $L_{38} {\sim}1.5$ (outer) and
$L_{38} {\sim}2.3$ (inner) in $10^{38}$ erg/sec units.

The nucleus and both arcs exhibit Ne enhanced relative to a solar
value.  While the arc values are higher, within the errors, they are
consistent with the nuclear value.  However, both arcs also exhibit
enhanced values for Mg of 2.5-3.5 times solar.  Additionally, the
inner arc has an enhanced value for oxygen of 3.3
(Table~\ref{fit_table1}).  That the arcs exhibit enhanced O and Mg
abundances suggests the outflow has swept up material from prior
phases of stellar mass loss.  The enhanced Ne found in the nucleus and
the arcs could indicate some nucleosynthetic processing occurred
before the gas was expelled.

\subsection{Multi-wavelength Observations}\label{OtherBands}

The X-ray observations are more easily interpreted if placed in
context with the emission from other wavelengths.  The position of the
NGC 5195 nucleus is sufficiently well-determined, in spite of its near
CCD-edge position, that it and the arc structures may be compared
with images at other wavebands.  The nuclear source corresponds within
a few arc seconds to the positions of the nuclear source in the data
sets from the {\it Swift} UVOT, DSS2 (blue), WISE 6${\mu}$, and VLA
FIRST images (Figure~\ref{tiled_field}).

Most other wavelengths reveal little structure at the location of the
arcs, particularly if the band is a continuum or has a wide wavelength
coverage.  Necessarily, that statement depends on the degree to
which the images in other bands provide sufficient dynamic range in
the area of the sky where the arcs are located (e.g., WISE 6${\mu}$
image has insufficient resolution to discern; DSS2 is over-exposed in
the same region).

However, three images are intriguing.  First, the 6 cm VLA image
(Figure~\ref{tiled_field}) shows a broad arc SE of the nucleus of NGC
5195 at approximately the distance of the outer X-ray arc.  The
breadth of the radio arc is likely tied to the resolution of the
configuration used (VLA configuration `D') for the observation.
Second, the {\it Swift} UVOT image (Figure~\ref{tiled_field}) exhibits
a structure to the north of NGC 5195 at a distance from the nucleus
comparable to the distance between the nucleus and the inner arc.  The
UVOT structure resembles a single broad arc, suggesting a possible
symmetric expulsion of matter.  That interpretation leaves a puzzle:
why are there two X-ray-visible arcs to the South and a single
UV-visible arc to the North?

The third interesting image is H${\alpha}$ (\citealt{Hoopes2001};
retrieved from the NASA Extragalactic Database;
Figures~\ref{tiled_field}, ~\ref{HalphaOverlaid}).  A slender
H${\alpha}$ arc and a second, less well-defined structure are visible
near the regions of the X-ray arcs.  The slender H${\alpha}$ arc lies
{\it outside} the outer X-ray arc, implying `snow-plowed,' or
swept-up material by the expanding plasma driven by the AGN outflow.
Based on a single spectrum obtained in 1998 with the spectrograph's
slit lying approximately across the H$_{\alpha}$ arc (their `slit 4'),
\cite{Hoopes2003} conclude that the spectral behavior is indicative of
an expanding structure or outflow, but that `the mechanism for such an
outflow is unclear.'  The same authors also note that this region
exhibits high [N~II]/H${\alpha}$, [S~II]/H${\alpha}$, and
[O~III]/H${\beta}$, with values higher than in the diffuse gas
elsewhere in the galaxy.  Those emission lines are indicative of
shocked emission.

{\bf The presence of the H${\alpha}$ emission plus the shock
indicators support an interpretation of the X-ray arcs as resulting
from shocks driving the H${\alpha}$ emission outward.  Earlier, we
raised the possibility that the arcs originate from collision-induced
star formation.  We interpret the shape of the H${\alpha}$ emission
and its location immediately outside the outer X-ray arc as support
for the outgoing blast wave interpretation, rather than {\it in situ}
collision-induced star formation activity.

Furthermore, collision-induced star formation often occurs in tidal
debris or along the bridges/tails linking the interacting pair
\citep{Machacek2009}.  The simulations to date either demonstrate a
glancing collision with NGC 5195 passing by on a hyperbolic trajectory
\citep{TT72}, or that there have been multiple encounters
with the two galaxies in high-eccentricity orbits \citep{SL2000}.  None of
the simulations we have seen to date have been done with sufficient
resolution to investigate detailed gas motions.}

The outer H${\alpha}$ arc is ${\approx}$1660 pc in length
(Table~\ref{arc_pos}).  Along the length of the arc, the width is
${\approx}5''$ or ${\approx}$200 parsecs.  Measured velocities ranged
from 500-600 km s$^{-1}$.  This arc also exhibits what appear to be
two or three small H~II regions, suggesting the outer arc has plowed
sufficient material to trigger star formation.

\section{Discussion}\label{disc}

The X-ray arcs represent a clear case of feedback from the SMBH: it
almost certainly cannot be a coincidence that the arcs arise in a
galaxy that is known to possess a SMBH and has undergone at least one
interaction with a nearby galaxy in the past ${\approx}10^8$ years.
In that, the observations presented here for NGC 5195 appear broadly
to match observations of outflows observed in other systems, as we
describe at the end of this section.

Large-scale AGN feedback plays essential roles in galaxy evolution. In
the initial quasar phase, there is very rapid growth of the central
black hole through radiatively efficient accretion, along with active
star formation. Nuclear winds driven by the quasar or, in the case of
ULIRGS, by starbursts, can significantly remove gas from galaxy cores,
thus quenching nuclear star formation, leading to radiatively
inefficient accretion (For a recent review of quasar feedback, see
KP2015.).  At the present epoch, central SMBHs are generally not in a
quasar mode, but instead are radiatively faint, releasing the energy
they produce from accretion kinetically through jets. These AGN
outbursts can reheat cooling gas, thus truncating star formation and
the growth of the SMBH.

AGN feedback is often observed through the cavities and shocks
produced by SMBH outbursts in the hot X-ray atmospheres of massive
early-type galaxies, groups and clusters. However the mass of NGC~5195,
estimated to be one tenth to half the mass of its companion NGC~5194
(\citealt{Smith1990}; \citealt{Howard1990}), is in the range from a
few $10^{10} M_{\odot} \dot{m}$ to $10^{11} M_{\odot} \dot{m}$ and is
too low to gravitationally bind hot X-ray emitting gas. Instead we
have argued that the X-ray and H$\alpha$ arcs near the nucleus of
NGC~5195 are likely produced by episodic outbursts from the central
SMBH that shock-heat cool gas in the galaxy core to X-ray emitting
temperatures.

For those less familiar with the literature on feedback, we direct
readers to three recent reviews, each with a slightly different view
of feedback: from the viewpoint of clusters and massive galaxies
(F12), from luminous AGN (KP2015), and from outflows caused by
starbursts in low-mass galaxies \citep{Erb2015}.

\subsection{Estimates of Physical Quantities}

The data at present are limited, so many of the estimates must be
order-of-magnitude.  There are also a number of quantities that
cross-feed into the estimates.  As we want to avoid creating confusion
in the reader's mind, the overall path is as follows: determine the
mass of the SMBH, then L$_{Edd}$, then an age estimate for the
outbursts from the velocities, followed by a density estimate, which
leads to the mass of the arcs, followed by an estimate of the cooling
time of the arcs.

From the velocity dispersion ${\sigma}$ of 124.8 km s$^{-1}$ for
NGC~5195 \citep{Ho2009}, the M$_{\rm BH}-{\sigma}$ relation yields
$M_{\rm BH} {\sim} 3.8{\times}10^7 M_{\odot}$ ($M_7 = 3.8$), assuming
${\sigma}^{\alpha} = {\sigma}^{4.4}$ (\citealt{FM2000},
\citealt{Gebhardt2000}).  The Eddington luminosity for NGC 5195 is
$$L_{\rm Edd} = 4 {\pi} G c M / {\kappa} {\approx} 1.3{\times}10^{45}
(M_{\rm BH}/10^7 M_{\odot})$$ where ${\kappa}$ = electron scattering
opacity; with values inserted for NGC 5195, we have $L_{\rm Edd} =
4.9{\times}10^{45}$ ergs s$^{-1}$ compared to the observed 0.5-2 keV
luminosity of ${\sim}5{\times}10^{38}$ ergs s$^{-1}$ (\S\ref{NucSpec}).
To reiterate, that value means the NGC 5195 SMBH is in a radiatively
inefficient state of $L_{obs} = f L_{Edd}$ with $f {\sim} 10^{-7}.$

A measure of the velocities of the arcs could lead to their estimated
age.  Unfortunately, the velocities of the arcs and their surroundings
likely have not been measured.  The velocity measured from the
\cite{Hoopes2001} H${\alpha}$ spectrum, ${\sim}500-600$ km s$^{-1}$,
was obtained from a long-slit spectrograph and represents an average
over a range of velocities (\S\ref{OtherBands}).  The velocity of NGC
5195 itself is ${\sim}$470 km s$^{-1}$ (NED).  With uncertainties
included, these two values are comparable.  Further, if the gas is
confined to the interaction plane, then any measured velocity is a
projection into our line-of-sight.  Absent a good measure of the
projection angle, measured velocities have a wide range.

Can limits on the velocity of the arcs be set or inferred?
\cite{Hoopes2001} measured line ratios from optical spectra from which
they inferred that the ratios are consistent with a shock velocity of
``several hundred kilometers per second."  Additionally, a lower limit
on the arcs' motion is set by considering the time from the
interaction of NGC 5195 and NGC 5194 based on models: ${\sim}$50-100
Myr \citep{SL2000}.  Given the present positions of the arcs, this
leads to velocities of ${\sim}8-15$ (inner) and ${\sim}17-35$ (outer)
km/sec.  However, velocities this low would {\it not} generate shock
emission consistent with the \cite{Hoopes2001} values.

If we instead assume that ``several hundred'' translates to
${\approx}$300 km/sec, then the arcs require ${\sim}2.5{\times}10^6$
(inner) and ${\sim}5.6{\times}10^6$ (outer) years to reach their
current positions and the episodes are separated by
${\approx}3{\times}10^6$ years.  Adopting the definition of terminal
velocity from \cite{King2010} (equation 37 in that paper), and
assuming this value is the velocity of the arcs, then with
${\sigma}_{200} = 0.624$, the terminal speed is $v_e {\sim} 875
{\sigma}_{200}^{2/3}$ km s$^{-1} {\approx} 620$ km s$^{-1}$ (We note
that the ${\approx}300-600$ km s$^{-1}$ range fits within the
simulations of \cite{SL2000}.).  This value leads to dynamical ages of
${\sim}1.2{\times}10^6$ (inner) and ${\sim}2.7{\times}10^6$ (outer)
years.  The episodes are then separated by ${\approx}1.5{\times}10^6$
years.  Consequently, if we assume the interaction actually occurred
${\approx}$50 MYr ago, then the observed feedback took some time to
develop, with the arcs emerging over the past ${\approx}10^{6-7}$ years.

{\bf One significant question then arises: what was the material doing
for ${\approx}$40 MYr, the time between the interaction of NGC 5195
with NGC 5194 and the outbursts?  The answer could be ``infalling'':
if we assume material started from rest and was located
${\approx}$1500 pc distant (approximately the current separation of
the arm of NGC 5194 from the nucleus of NGC 5195), then ${\approx}$65
MYr must elapse before the gas reaches the nucleus.  With an arbitrary
initial speed of 100 km s$^{-1}$, that time would be reduced to
${\approx}$15 MYr.  Absent additional constraints, this estimate may
be the best we can do.}

We now move to estimating gas densities with the goal of calculating
masses and cooling times for the arcs to compare with the dynamical
age estimate.  The H$_{\alpha}$ (Figure~\ref{HalphaOverlaid}) arc
provides one route to the matter density.  The flux in the line
emission, integrated over a rectangular region matching the
H$_{\alpha}$ arc, and again adopting a thickness = 100 $thick_{100}$
pc, leads to a volume V of 1.3 $thick_{100}$ ${\times}10^{63}$ cm$^3$.
Assuming the Lyman H lines are optically thick (`Case B'
recombination) then $L = n {\Lambda} V$ and adopting an emissivity
${\Lambda}$ of 2.59${\times}10^{-13}$ cm$^3$ s$^{-1}$ (\citealt{OF},
table 2.1), leads to a number density of ${\sim}5/thick_{100}$
cm$^{-3}$ or a mass estimate of ${\approx}4{\times}10^6$ M$_{\odot}$.
That value is mid-range for typical interstellar molecular clouds
\citep{Murray2011}.

A second estimate of the number density in the X-ray arcs follows from
the X-ray {\tt vapec} model normalization:
$$normalization~=~{{10^{-14}}\over{(4{\pi}(D_A (1 +
z))^2}}~{\int}~n_e~n_p~dV$$ \citep{Arnaud1996}.  Assuming $n_e = n_p,$
z = 0.002, D$_A$ = 8 d$_8$ Mpc, and using the volume V as defined in
the prior paragraph, then $n_e {\approx} 0.06(0.09) thick^{-1}_{100}$
cm$^{-3}$ for the outer (inner) arc.  The mass in each of the
X-ray-emitting arcs is then ${\approx}5.5{\times}10^4 M_{\odot}$ (5.7
(outer) and 5.3 (inner)) or about 1\% of the H${\alpha}$ arc.  

Given the mass disparity, can the X-ray arcs be pushing the
H${\alpha}$ arc?  Assuming the gas in the X-ray and the H${\alpha}$
arcs may be described by the ideal gas law, then the pressures for the
H${\alpha}$, outer, and inner X-ray arcs are $3.5{\times}10^{-12}$,
$1.9{\times}10^{-11}$, and $4.3{\times}10^{-11}$ dyn cm$^{-2}$,
respectively.  The X-ray arcs then have about a factor of 5-12 times
higher pressure than the H${\alpha}$ arc, driving material outwards.
The total kinetic energy of the X-ray arcs is
${\approx}2{\times}10^{53}$ ergs.

A second check on our mass estimates is possible.  If the outer arc
swept clear the gas in the halo along that vector, then the mass in
the second arc is an approximate measure of the rate of accumulation.
We acknowledge that the second arc is not moving through precisely the
same space, but the question remains: is it possible to sweep up the
observed mass?  Above, we estimated the mass in the second arc as
${\approx}5.3{\times}10^4$ M$_{\odot}$ and the interval between
episodes as ${\approx}1.3{\times}10^6$ years.  Those values lead to an
accumulation rate of ${\approx}4{\times}10^{-2}$ M$_{\odot}$
yr$^{-1}$.  \cite{Magorrian1998} developed a bulge mass-SMBH mass
relation with updates by \cite{McConnell2013} and \cite{Belfiori2012}
and environmental dependencies explored by \cite{McGee2013}.  The
value of ${\sigma}_{100}$ and the SMBH mass yield an estimated bulge
mass / SMBH mass ratio of ${\approx}$1000, or a bulge mass of
${\approx}2-4{\times}10^{10}$ M$_{\odot}$\footnote{We note that this
relation has a two orders-of-magnitude range for any given SMBH mass,
but the range is not critical for our purposes: our intent in using
the relation at this time is plausibility, not accuracy.}.  If we
assume most of that mass is composed of ${\leq}1 M_{\odot}$ stars
given the color of NGC 5195, then some of those stars will have
evolved off the main sequence to at least the asymptotic giant branch
(AGB).  AGB stars have typical mass loss rates of ${\approx}10^{-8}$
M$_{\odot}$ yr$^{-1}$ \citep{DeBeck2010}.  If just 0.1\% are AGB
stars, then the mass available per year is ${\approx}10^{-1}$
M$_{\odot}$, more than sufficient to provide the swept-up mass
observed.  As a by-product, we would expect sector-to-sector
differences in gas density and metals in these regions, a potential
test of the overall interpretation described in this paper.

With densities in hand, we can estimate the cooling time for
comparison with the dynamical time.  We use Figure~1 of
\cite{Gehrels1993} which displays, for an optically-thin gas with
solar abundances, the cooling curve as a function of temperature for
several processes (two-photon, recombination, bremsstrahlung, and line
emission).  For the measured temperatures of the arcs of ${\sim}$0.4
and 0.65 keV, line emission dominates the cooling (the next largest
contributor, bremsstrahlung emission, contributes ${\leq}10\%$).  The
cooling time is defined as $$t_{cool} = (3/2) kT / (n_e
{\Lambda}(T))$$ {\noindent}where ${\Lambda}(T)$ is the cooling
function.  With values inserted, we obtain an estimated cooling time
of ${\approx}1-1.6{\times}10^7$ yr for the X-ray arcs.  This estimated
cooling time is about an order-of-magnitude larger than the dynamical
time.  We then expect X-ray spectra should be dominated by line
emission, a prediction that could be checked using {\it Hitomi}
\citep{Takahashi2014}.

To summarize, we can place a crude limit on the age of the arcs
(${\sim}1-3{\times}10^6$ yrs) that is an order-of-magnitude lower than
the cooling time for the arcs based on crude estimates of their
velocities.  Estimated densities (${\approx}0.1-5$ cm$^{-3}$) lead to
the masses of the arcs, which combined with the arcs' velocities
(adopted as ${\sim}$600 km s$^{-1}$), carry sufficient momentum to
sweep up the H${\alpha}$ arc.  The estimated densities are supported
by an estimate of the normal stellar mass loss within the galaxy's
nuclear volume over the lifetime of the arcs.

\subsection{Additional Future Constraints on Source Parameters?}

\subsubsection{Transition Radius from Momentum- to Energy-driven Flow?}

With regard to AGN feedback, we can estimate the radius of transition
from momentum- to energy-driven flow (KP2015): $$R_{\rm crit} {\sim}
500~M_8^{1/2}~{\sigma}_{200}~{\rm pc}.$$ {\noindent}Assuming their
approach is valid for NGC 5195, the critical radius is ${\approx}$230
pc or ${\approx}5.8''$ at the adopted distance of M51.  For
comparison, the sphere of influence for the SMBH is $$R_{\rm inf}
{\sim} {{3 M_7}\over{{\sigma}_{100}^2}} {\sim}9 {\rm pc}
{\approx}0''.2.$$ {\noindent} The X-ray arcs both lie outside of this
critical radius and, within the context of KP2015, both arcs would
fall into the energy-driven regime.  $R_{\rm crit}$ could easily be
constrained with current-generation telescopes provided the extreme
luminosity contrast of the nucleus {\it vs} its surroundings is
overcome.

\subsubsection{Another Burst?}

Is there evidence of a third episode of feedback?  If high-resolution
spectroscopy of the nucleus of NGC 5195 reveals high-velocity
absorption lines, then a third episode would be underway.  We would
expect any possible third episode to be weaker in the sense of lower
mass in the outflowing material {\bf because the time interval between
the current burst is less than the time between the two outbursts.}

\subsubsection{Star Formation?}

Inspecting the H$_{\alpha}$ image reveals two or three knots of
emission (depending upon one's definition of 'knot'), strongly
suggesting the development, perhaps the {\it initial} development, of
H~II regions surrounding star formation.  Given the pressure and
densities involved, the expanding X-ray arcs should create the
conditions for star formation through the Rayleigh-Taylor instability.
If confirmed, this region in NGC 5195 would be a clear case of
triggered star formation.  Verification that stars are forming in the
plowed material will require additional optical or infrared
observations centered on the knots.  Estimates of the ages of the
star-forming complexes would aid in constraining the scenario
described in this paper.  In addition, careful inspection of the
sectors along which the arcs have travelled in comparison to other
sectors may provide a measure of the degree to which cold gas has been
removed by the action of the expanding arcs, thereby establishing at
least one timeline for galaxy evolution.

\subsubsection{Azimuthally-limited Arcs?}

That each arc is limited azimuthally may suggest an interpretation
that the arcs are more similar to a bubble.  That interpretation could
also argue for the North-South differences: bubbles present on the
South side of the nucleus need not also appear on the North side.
However, a bubble implies a slow vertical rise, hence unlikely to lead
a shock or to X-ray emission\footnote{Bubbles do have X-ray-emitting
rims because as they rise, they can sweep up {\it hot} gas in a
galaxy's halo \citep{Finoguenov2008}.  The NGC 5195 halo does not appear
to be hot, hence we would not expect to see hot, rimmed bubbles.}

Alternatively, \cite{King2006} describe a `chaotic accretion' scenario
in which the orientation of the accretion disk changes after an
accretion episode.  This scenario is used to describe SMBH growth at
high redshift.  Perhaps NGC 5195 is a local analog of events that took
place at high redshift.  The chaotic accretion scenario leads to
momentum feedback that is isotropic over long times.  That suggests
that over short times, feedback effects need not be isotropic.  While
we do not have any constraints on the orientation of an accretion disk
in NGC 5195, a question immediately arises from the observed arcs --
do their positions and separations constrain our understanding of the
accretion episodes or the interactions of the galaxies in any
fundamental way?  At this moment, we do not have an answer, but such a
question calls for additional simulations of the interactions.

\subsubsection{Similar Objects?}

Are there similar objects to NGC 5195?  Locally, there appear to be
very few -- one recent object may be NGC 660 \citep{Argo2015}.  It is
described as an `unremarkable starburst' with a `polar-ring morphology
and a LINER-type nucleus.'  It has an estimated SMBH mass of log
M$_{\rm BH} = 7.35 M_{\odot}$ and an Eddington ratio of $10^{-6}$.
Both values are similar to the NGC 5195 values.  Interestingly,
\cite{Argo2015} infer that a flare occurred between 2008 and 2012.  We
suggest NGC 5195 lies at an intermediate stage between a flare in the
SMBH and rings and bubbles observed in the outskirts of an AGN or
intercluster medium.  That the arcs originate with a SMBH, are
cooling, and appear to be moving at high velocity, raises a possible
third scenario\footnote{Instead of accretion from outside the Milky
Way, or supernova-driven fountains \citep{Fraternali2015}.} that some of the
high-velocity clouds observed in the Milky Way's halo \citep{Wakker1997}
originate in a similar manner.

How do the NGC 5195 arcs fit into the overall developing picture of
feedback?  Several objects studied with {\it Chandra} have been shown
to illustrate bubbles and outflows: arm-like shocked outflows in the
halo of NGC 4636 \citep{Jones2002}; bubbles near the center of M84
\citep{Finoguenov2008}; outflows in the E galaxy NGC 4552 (M89) in the
Virgo cluster \citep{Machacek2006} and in the center of the NGC 5846
group \citep{Machacek2011}.  The difference between NGC 5195 and these
objects is perhaps the difference between a relatively steady influx
of matter versus the transient influx to the SMBH in NGC 5195
resulting from the interaction with NGC 5194.  That comment suggests
that other interacting galaxies should exhibit similar reactions
provided the interaction has occurred relatively recently.

\section{Summary}

From X-ray and H$_{\alpha}$ images, we find two X-ray arcs on the
south side ${\sim}15''-30''$ from the nucleus of NGC 5195;
H$_{\alpha}$ arc lies just outside of the outer X-ray arc.  The
results for NGC 5195 are important for at least three reasons: (i)
they demonstrate dynamical and active feedback; (ii) they may
illustrate an example of triggered star formation; and (iii) the arcs
are a `snapshot' intermediate in the evolution between winds detected
in the X-ray spectra of nuclei of AGN and the large-scale filaments
and bubbles observed in the hot atmospheres of many massive galaxies
as well as in groups and clusters.

NGC 5195 is important for another reason: it is an example, perhaps
the only one detected to date, where at least four critical clocks are
evolving simultaneously: one clock started with the interaction
between NGC 5194 and NGC 5195; a second started with the star
formation triggered by the interaction; a third started with the
reaction of the SMBH to the accreted material from the interaction;
and the last clock commenced with the episodic nature of the response
of the SMBH -- given an accretion pulse, how long before the response?
The rates at which the clocks advance are not well-known, but NGC 5195
presents perhaps the nearest opportunity to at least constrain some or
all of those rates.  Knowledge of such rates may be important for
studies of the early universe.

Additional observations of NGC 5195 are necessary including: (i) a
deep {\it Chandra} observation north of NGC 5195 will ascertain the
outflow symmetry to test the expectation of isotropic outflows; (ii)
H${\alpha}$ velocities of both of the arcs and their surroundings,
using an integral field spectrometer, will pin down motions of the
blast waves and provide improved estimates of the mass of swept-up
material; (iii) estimates of the ages of star clusters that appear to
be forming in the swept-up material will further constrain the
timeline; (iv) observations at wavelengths that peer through the dust
may provide additional insights interior to the arcs, e.g., molecular
outflows; (v) observations that look {\it at} the dust can determine
whether it is optically thick (as noted in KP2015); (vi) polarization
observations that resolve field behavior may illuminate the
interaction of the magnetic field with the blast waves; and (vii) VLA
observations using the A configuration, or VLBI, can study the radio
emission in greater detail.  Given the dynamic nature of the arcs, 
high-resolution observations should be repeated on a moderate time
frame to probe possible time dependence.

\acknowledgements

We thank the referee for comments that improved our paper.  The
research of EMS was partially supported by the Vaughan Family
Endowment.

\clearpage

\begin{table}
\centering
\caption{List of M51 Observations}
\label{m51_obs}
\begin{tabular}{rlr}
        & Observation & Observed   \\
 ObsID  &  Date       &  ExpT (ks)  \\ \hline
\multicolumn{3}{c}{Faint mode} \\
~~354 & 2000 Jun 20 & ~14.9 \\
13812 & 2012 Sep 12 & 157.5 \\
13813 & 2012 Sep ~9 & 179.2 \\
13814 & 2012 Sep 20 & 189.8 \\
13815 & 2012 Sep 23 & ~67.2 \\
13816 & 2012 Sep 26 & ~73.1 \\
15496 & 2012 Sep 19 & ~40.9 \\
15553 & 2012 Oct 10 & ~37.6 \\
\multicolumn{3}{c}{Very Faint mode} \\
~1622 & 2001 Jun 23 & ~26.8 \\
~3932 & 2003 Aug ~7 & ~47.9 \\
12668 & 2011 July 3 & ~~9.9 \\
\hline
\end{tabular}

See \S\ref{DataPrep} for a discussion of the ObsIDs used.
\end{table}

\begin{table}
\begin{center}
\caption{Approximate Spatial Descriptions of the Arcs}
\label{arc_pos}
\begin{tabular}{lrrrr}
                         & \multicolumn{2}{c}{X-ray}   &  \multicolumn{2}{c}{H$_{\alpha}$} \\ 
Description              & Inner Arc & Outer Arc & Inner Arc & Outer Arc   \\ \hline
~~Mid-arc RA, Dec (J2000) & \\
~~~ 13:29:00, 47:15:00 +  & 58.5, 43  & 59, 22    & 58.5, 43  & 59.7, 13.5 \\
~~Overall length-arcsec   & 20        & 28        & 22        & 43   \\
~~Overall length-parsecs  & 775       & 1080      & 850       & 1670 \\
~~Width-arcsec            & 9         & 9         & 9         & 6    \\
~~Width-parsecs           & 350       & 350       & 350       & 230  \\
From nucleus:            \\
~~Outermost point-arcsec  & 20        & 44        & 29        & 48   \\
~~Outermost point-parsecs & 780       & 1700      & 1120      & 1700 \\
~~Innermost point-arcsec  & 13        & 31        & 15        & 41   \\
~~Innermost point-parsecs & 500       & 1200      & 580       & 1590 \\
\hline
\end{tabular}
\end{center}
Note: Arc positions are defined by rectangles enclosing $>$80\% of the
emission.  Position of nucleus adopted as (J2000) 13:29:59.577,
47:15:58.39 (NASA Extragalactic Database).  One arc second at M51
${\sim}$38.8 parsecs.  All values in parsecs should be multiplied by
d$_8$.
\end{table}

\begin{table}
\begin{center}
\caption{Spectral Model Fits: Nucleus and Arc Emission}
\label{fit_table1}
\begin{tabular}{lrrr}
       Parameter           &     Nucleus               &    Inner Arc             &  Outer Arc             \\  \hline
${\chi}^2$/dof; dof        & 1.31; 490                 & 1.56; 210                & 1.91; 228              \\
kT (keV)                   & 0.61$_{-0.05}^{+0.06}$    & 0.65$_{-0.04}^{+0.05}$   & 0.38$_{-0.03}^{+0.06}$ \\
N$_{\rm H}$ (10$^{22}$ cm$^{-2}$) & 0.15$_{-0.05}^{+0.06}$  & 0.021f$^a$          & 0.021f$^a$             \\
norm (${\times}10^{-6}$)   & 34.5$_{-1.27}^{+1.47}$    & 5.02$_{-0.53}^{+0.55}$   & 7.97$_{-0.59}^{+0.64}$ \\
Abundance-O ($Z_{\odot}$)  & 1.0f                      & 3.0$_{-0.9}^{+1.2}$      & 1.6$_{-0.5}^{+0.7}$    \\
Abundance-Ne ($Z_{\odot}$) & 3.61$_{-0.90}^{+1.08}$    & 4.1$_{-1.0}^{+1.2}$      & 3.7$_{-1.0}^{+1.5}$    \\
Abundance-Mg ($Z_{\odot}$) & 1.0f                      & 1.9$_{-0.5}^{+0.7}$      & 2.8$_{-0.9}^{+1.8}$    \\
 \\
Powerlaw                   & 1.17$_{-0.38}^{+0.31}$    & 1.6f                     & 1.6f                   \\
~~norm (${\times}10^{-7}$) & 6.71$_{-0.21}^{+0.25}$    & $<$7.0                   & $<$5.0                 \\
\\
Background (used with all components)  &    \\
Index                      & 0.24$^{+0.18}_{-0.25}$ & ${\cdots}$ & ${\cdots}$\\
Index                      & 4.78$^{+1.50}_{-0.98}$ & ${\cdots}$ & ${\cdots}$\\
norms (${\times}10^{-7}$):   \\
~~~power law              & 2.64$^{+0.52}_{-0.60}$  & ${\cdots}$ & ${\cdots}$ \\
~~~power law              & 0.82$^{+0.85}_{-0.52}$  & ${\cdots}$ & ${\cdots}$ \\
~~~gaussian-1: 0.567 keV  & 1.56$^{+0.77}_{-0.83}$  & ${\cdots}$ & ${\cdots}$ \\
~~~gaussian-2: 0.764 keV  & 0.62$^{+0.24}_{-0.29}$  & ${\cdots}$ & ${\cdots}$ \\
~~~gaussian-3: 1.767 keV  & 0.43$^{+0.11}_{-0.10}$  & ${\cdots}$ & ${\cdots}$ \\
~~~gaussian-4: 2.129 keV  & 0.63$^{+0.17}_{-0.15}$  & ${\cdots}$ & ${\cdots}$ \\
\\
Measured Flux (${\times}10^{-14}$, 0.5-2 keV) & \\
~~Source                   & 4.42${\pm}$0.40           & 3.06${\pm}$0.28          & 1.97${\pm}$0.17        \\
~~Background               & 0.11                      & 0.14                     & 0.18                   \\ 
Luminosities (${\times}10^{38}$)  \\
~~Source (absorbed)        & 3.37                      & 2.33                     & 1.50                   \\
~~Source (unabsorbed)      & 3.37                      & 2.52                     & 1.62                   \\
~~Background               & 0.08                      & 0.11                     & 0.14                   \\
\hline
\end{tabular}
\end{center}

Table notes: The `f' on any parameter indicates it was fixed at the
value listed$^a$(for the column density, see \S\ref{arcspectra} for
discussion).  For the power law index of the background, a fit
determined the value, then it was fixed.  The energy range used in the
spectral fits was 0.4 - 2.4 keV.  Error values represent the 90\%
contour.

\end{table}

\clearpage

\setcounter{figure}{0}

\begin{figure}
\centering
\caption{(b) an expanded view of (a) with the two arcs south of the
nucleus, as well as the nucleus, indicated.  The image is un-binned
but zoomed by a factor of 2.  The other dark objects are X-ray point
sources either near the nucleus or in the background.  The dashed line
indicates the approximate edge of the CCD.  The black-and-white inset
shows four (red arcs) of the twenty regions used to determine the
radial profile.}
\scalebox{0.6}{\includegraphics{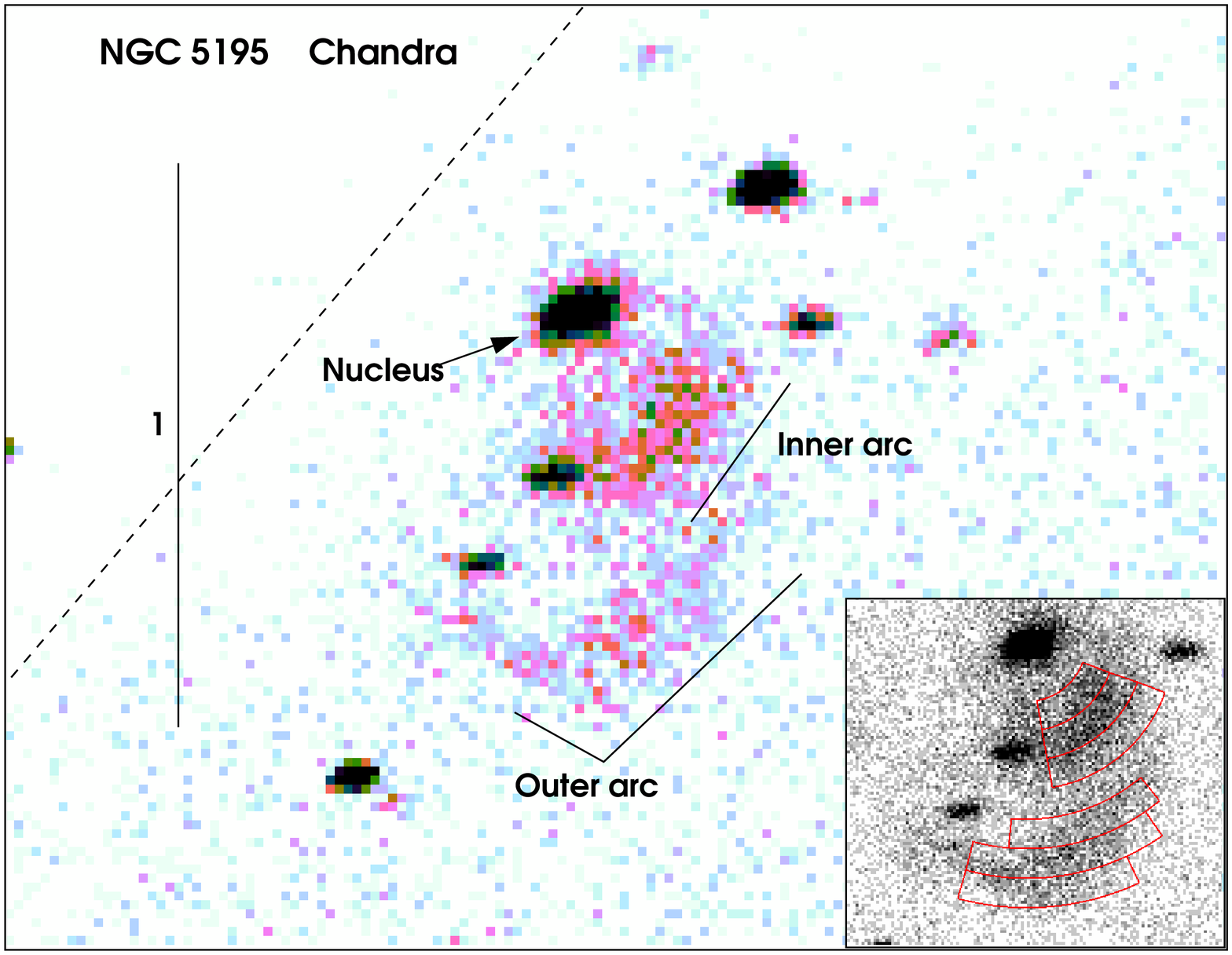}} 
\end{figure}

\setcounter{figure}{0}

\begin{figure}
\centering
\caption{(c) Comparison of the summed ${\sim}$700 ksec Faint mode data
(right) with the ${\sim}$790 ksec merged Faint + Very Faint mode data
(left).  North is up, East is left.  Absent the longer exposure, the
Very Faint mode data would be difficult to interpret.  However, in the
context of the long Faint mode exposures, the faint emission North of
the nucleus in this image may suggest symmetric expulsion of matter.
The dashed line in the right image indicates the approximate edge of
the CCD.}
\scalebox{0.4}{\includegraphics{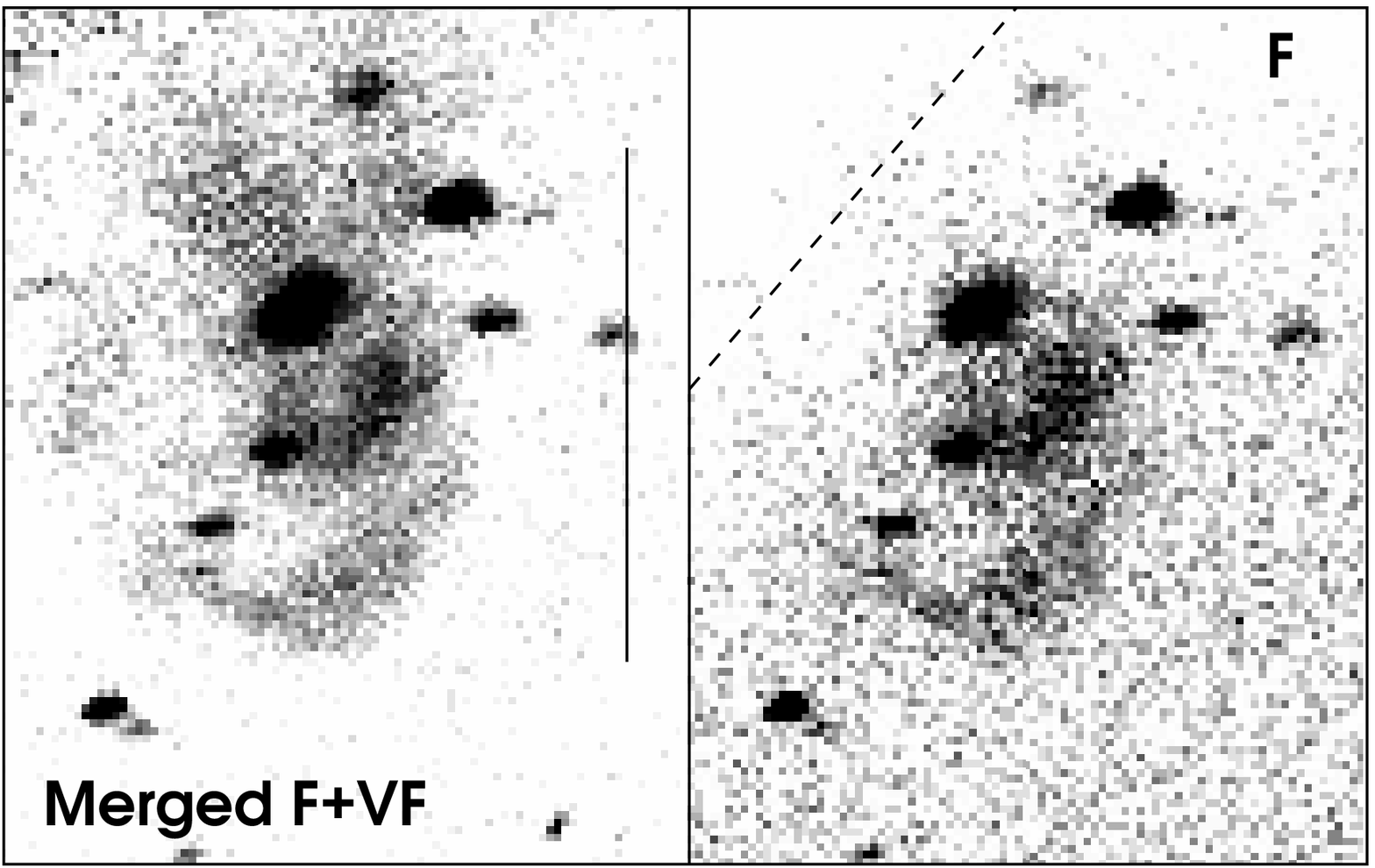}}
\end{figure}

\begin{figure}
\centering
\caption{Radial profiles of both arcs.  The counts were extracted in
angular annuli using a region similar to that shown in
Figure~\ref{n5195_structure}(b) but using 20 annuli.  The radius is
the mid-point of an annulus and is defined as the radial distance from
the nucleus.  The position of the nucleus is adopted as defined in
Table~\ref{arc_pos}. The data were at full resolution, so each pixel
is $0''.5{\times}0''.5$.  The inner arc partially overlaps the outer
arc near position angle ${\sim}$180 degrees; the background region for
the inner arc is consequently shorter in azimuth than the region for
the outer arc.}
\label{RadProf}
\scalebox{0.6}{\includegraphics{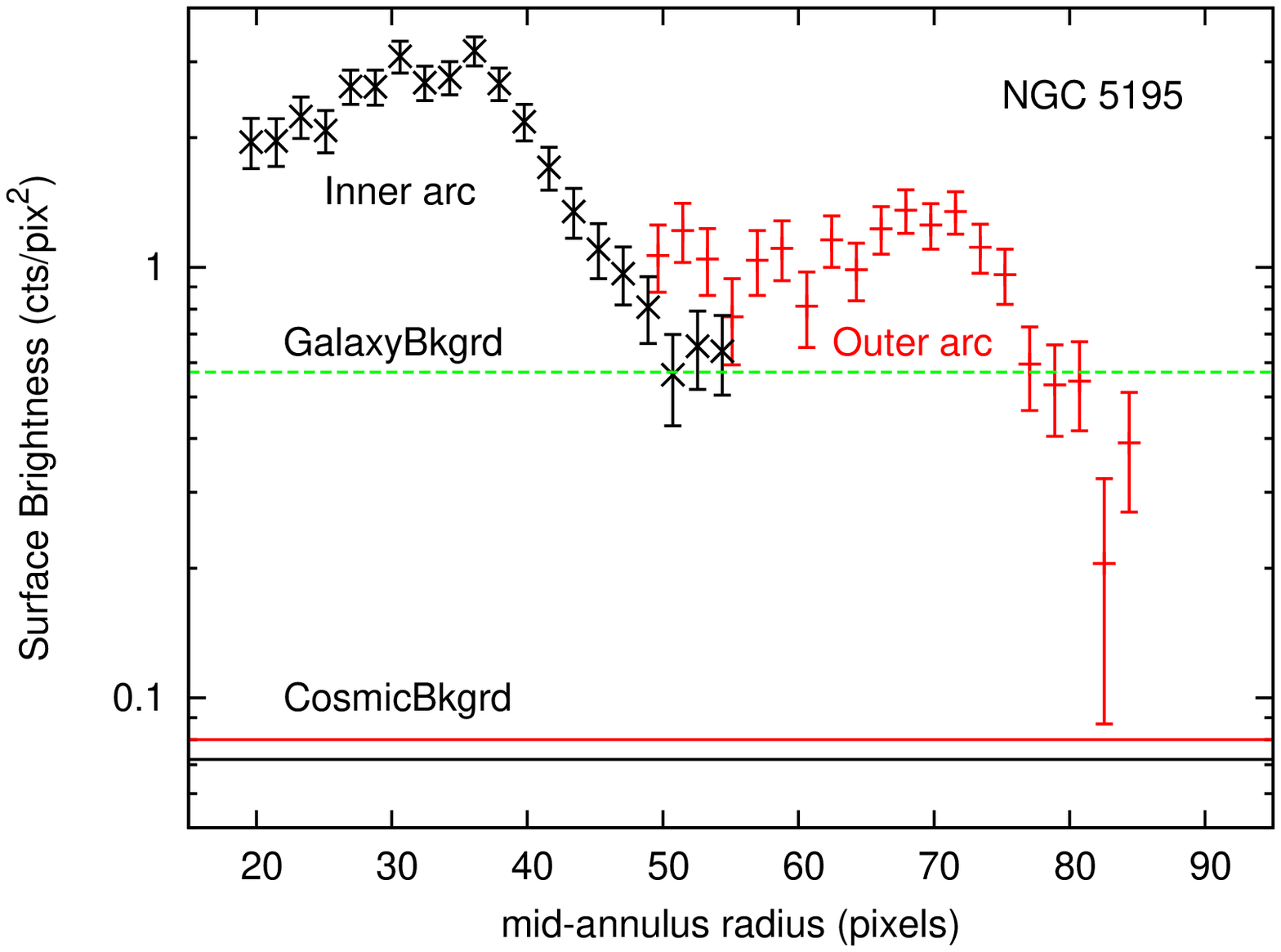}}
\end{figure}

\begin{figure*}
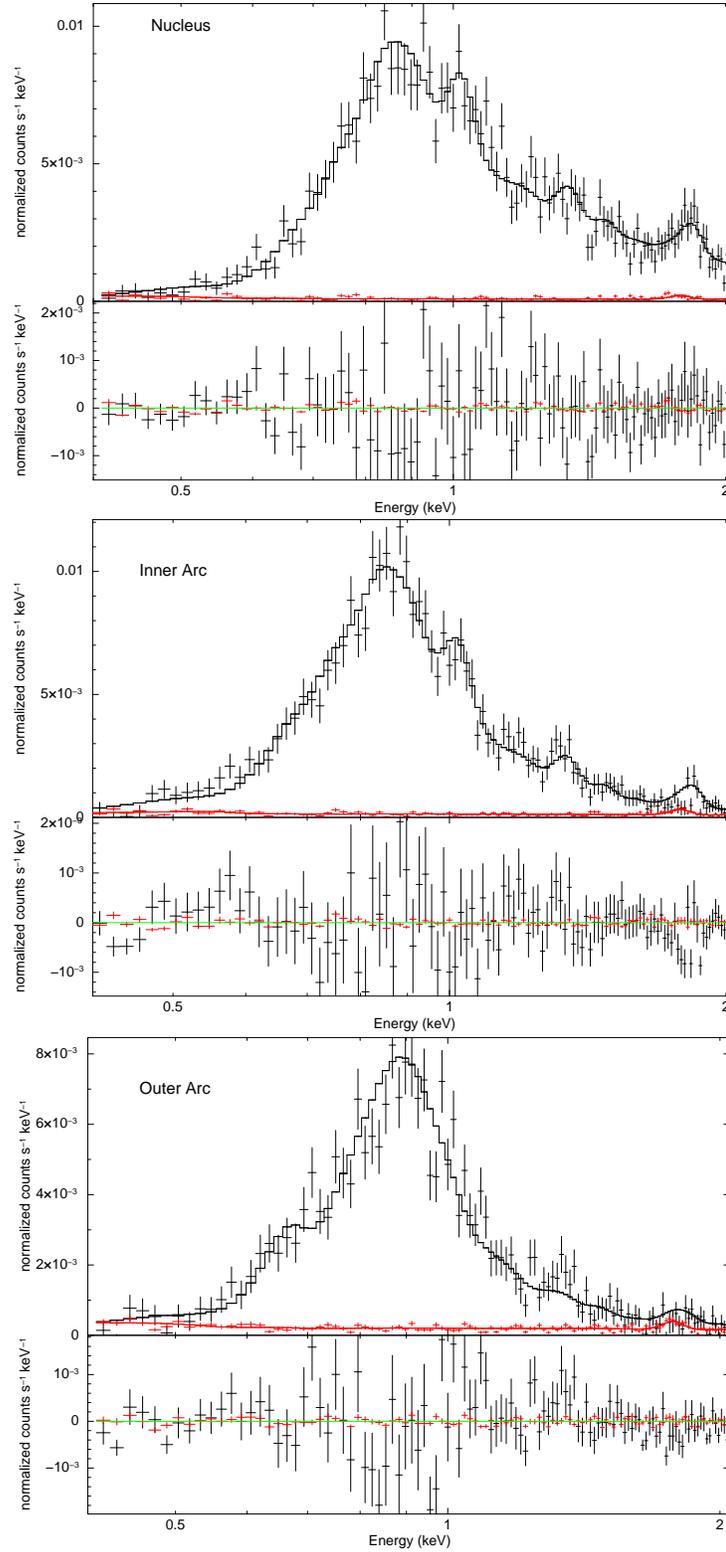
 
\centering 
\caption{Spectra of the (a) nucleus and (b, c) the inner and outer
diffuse arcs.  Note the differences in the spectra, particularly the
continuum slopes and bumps at energies of ${\sim}$1 (Ne) and 1.4 (Mg).}
\label{n5195_spectra}
\scalebox{0.4}{\rotatebox{-90}{\includegraphics{Schlegel_N5195_F3a.eps}}}
\scalebox{0.4}{\rotatebox{-90}{\includegraphics{Schlegel_N5195_F3b.eps}}}
\scalebox{0.4}{\rotatebox{-90}{\includegraphics{Schlegel_N5195_F3c.eps}}}
\end{figure*}

\begin{figure*}
\centering
\caption{Multi-wavelength view of the NGC 5195 structure: {\it
Chandra} X-ray; {\it Swift} UVOT (uw1); DSS2 Blue; H${\alpha}$; {\it
WISE} 6${\mu}$; VLA 6 cm.  North is up and East is left in all frames.
The size of each frame is identical and ${\sim}54$ arc sec on a side.
The contours represent the X-ray diffuse arcs with contours drawn at
10, 12.5, 15, 17.5, and 20 counts.  The H${\alpha}$ frame is expanded
in Figure~\ref{HalphaOverlaid}.  Note that the outer X-ray arc
overlays a broad arc-like feature in the VLA 6 cm observation.}
\label{tiled_field}
\scalebox{0.5}{\includegraphics{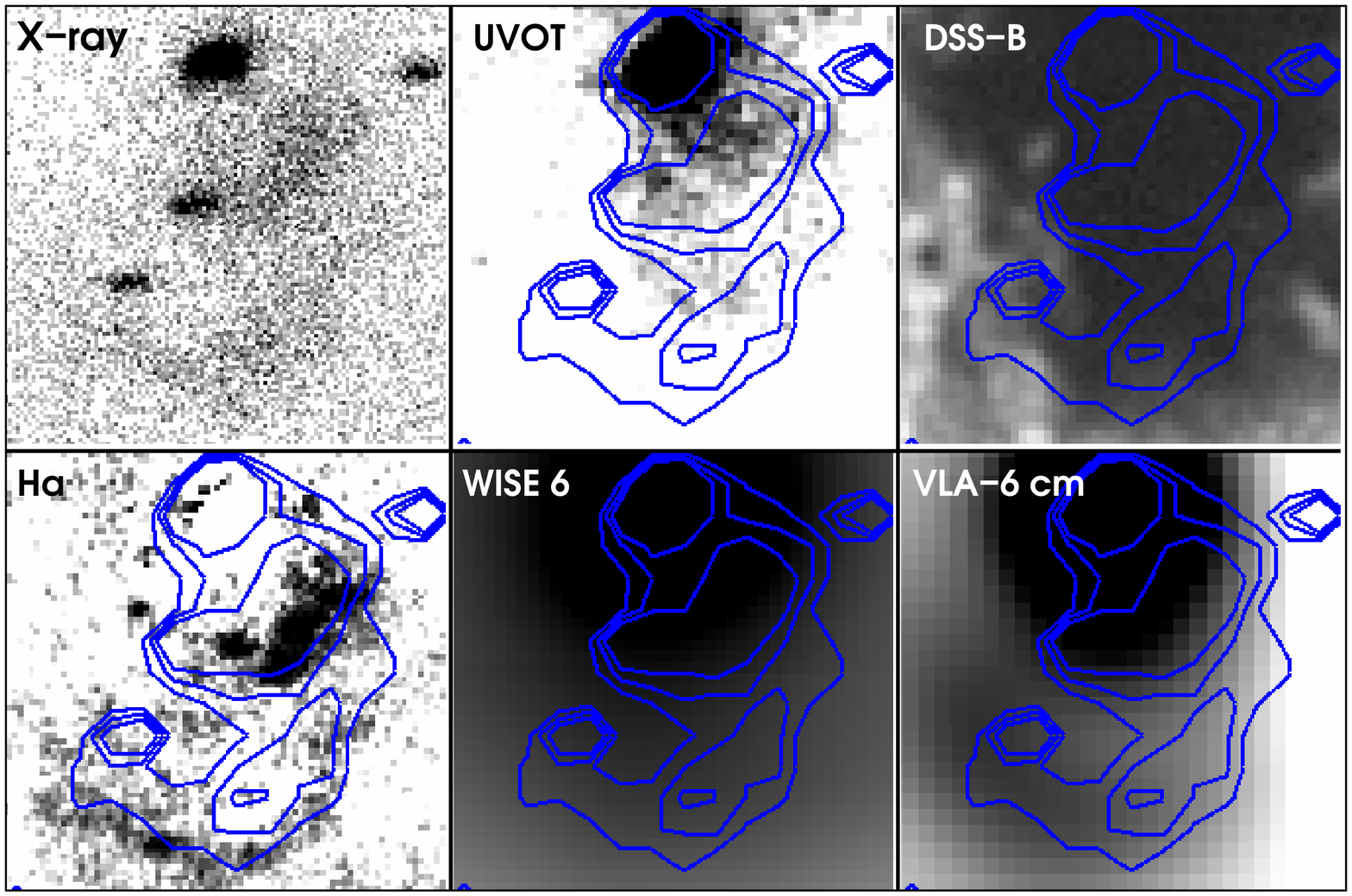}}
\end{figure*}

\begin{figure*}
\centering
%\caption{(a) The H${\alpha}$ image of NGC 5195 with contours of X-ray
\caption{The H${\alpha}$ image of NGC 5195 with contours of X-ray
emission (blue) overlaid from Figure~\ref{tiled_field}.  The
contours are as described in Figure~\ref{tiled_field}.  North is up
and East is left.  The scale bar on the right side is 1 arc minute in
length.  The short bars approximately in the middle of the upper and
left sides of the box point toward the nucleus of NGC~5195, indicated
by a small, red cross.  The line of H${\alpha}$ lumps directly east of
NGC 5195 are H~II regions on the end of the arm of NGC 5194, the
nucleus of which lies nearly due south of NGC 5195.}
\label{HalphaOverlaid}
\scalebox{0.6}{\includegraphics{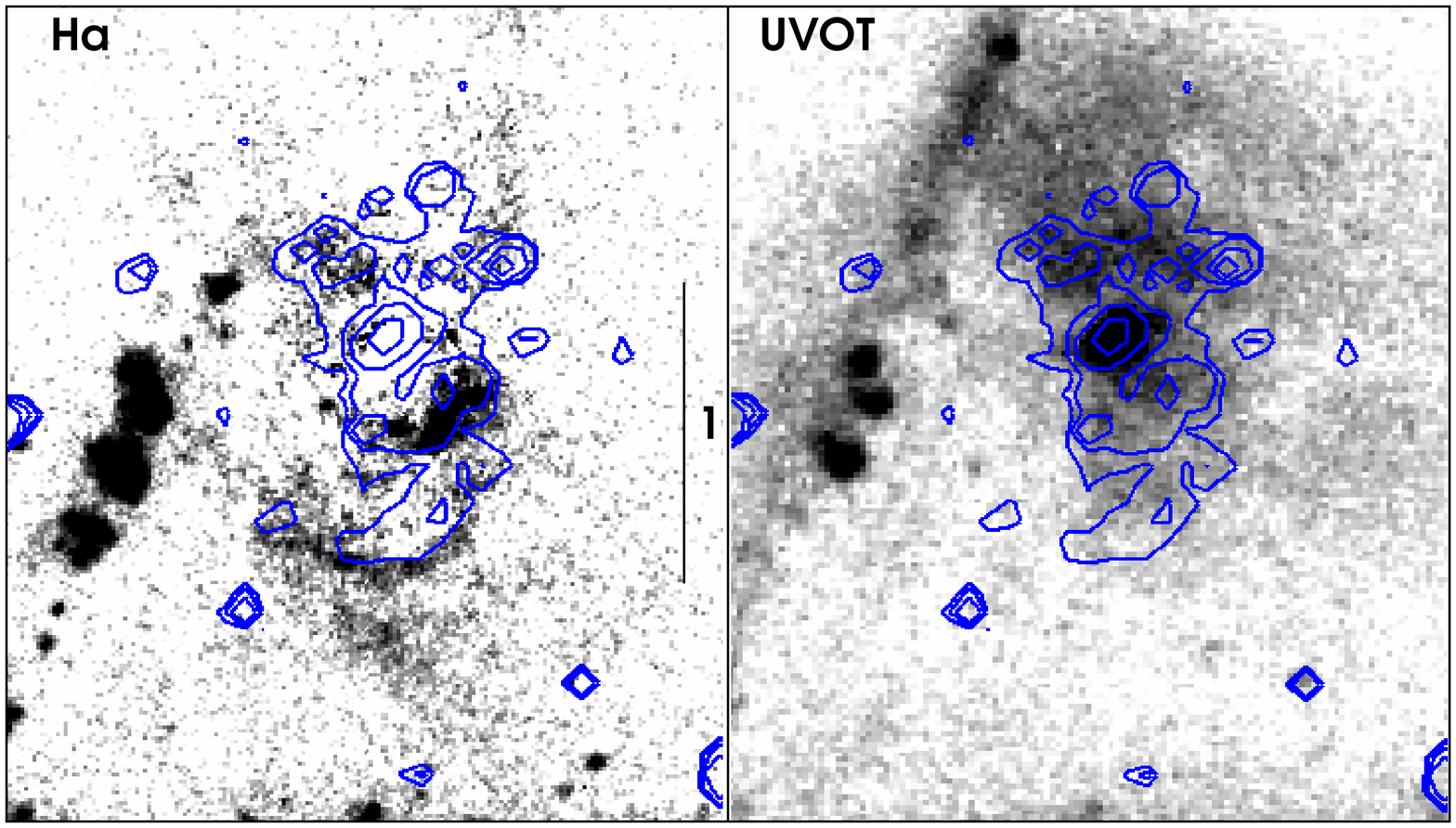}}
\end{figure*}

\end{document}